\documentclass[prd,twocolumn,preprintnumbers,superscriptaddress,nofootinbib]{revtex4-1}
\pdfoutput=1
\usepackage[a4paper, hdivide={1.91cm,,1.165cm}, vdivide={1.83cm,,3.0cm}]{geometry}

\usepackage{amsmath}
\usepackage{graphicx}
\usepackage{xspace}
\usepackage{color}
\usepackage{units}
\usepackage{slashed}
\usepackage[hyperfootnotes=false]{hyperref}
\usepackage{gensymb}

\def \L {\mathcal{L}} %Lagrangian density

\def \epsilon {\varepsilon} %different epsilon symbol

\newcommand{\matrixx}[1]{\begin{pmatrix} #1 \end{pmatrix}} %Matrix with brackets

\newcommand{\hc}{\ensuremath{\text{h.c.}}}
\newcommand{\Br}{\text{Br}}
\newcommand{\tr}{\text{tr}}
\newcommand{\diag}{\text{diag}}
\def \i {\mathrm{i}\mkern1mu} %imaginary unit i to distinguish it from indices i,j,etc.

\newcommand*{\id}{{1\hspace{-0.87ex}1}} % identity matrix without having to load packages

\allowdisplaybreaks
\begin{document}
%%%%%%%%%%%%%%%%%%%%%%%%%%%%%%%%%

\title{ The Majoron at two loops
}

\preprint{UCI-TR-2019-23, \href{http://arxiv.org/abs/1909.02029}{arXiv:1909.02029}, \href{https://doi.org/10.1103/PhysRevD.100.095015}{Phys.~Rev.~\textbf{D100} (2019) 095015}}

\author{Julian Heeck}
\email{Julian.Heeck@uci.edu}
\affiliation{Department of Physics and Astronomy, University of California, Irvine, CA 92697-4575, USA}

\author{Hiren H. Patel}
\email{hpatel6@ucsc.edu}
\affiliation{Department of Physics and Santa Cruz Institute for Particle Physics,
University of California, Santa Cruz, CA 95064, USA}

\hypersetup{
    pdftitle={The Majoron at two loops},
    pdfauthor={Julian Heeck, Hiren H. Patel}
}

%%%%%%%%%%%%%%%%%%%%%%%%%%%%%%%%%%
%%%%%%%%%%%%%%%%%%%%%%%%%%%%%%%%%%

\begin{abstract}

We present singlet-Majoron couplings to Standard Model particles through two loops at leading order in the seesaw expansion, including couplings to gauge bosons as well as flavor-changing quark interactions.
We discuss and compare the relevant phenomenological constraints on Majoron production as well as decaying Majoron dark matter.
A comparison with standard seesaw observables in low-scale settings highlights the importance of searches for lepton-flavor-violating two-body decays $\ell \to \ell' +$Majoron in both the muon and tau sectors.

\end{abstract}

%%%%%%%%%%%%%%%%%%%%%%%%%%%%%%%%%
%%%%%%%%%%%%%%%%%%%%%%%%%%%%%%%%%
\maketitle

%%%%%%%%%%%

\tableofcontents

\section{Introduction}

The Standard Model (SM) has emerged as an incredibly accurate description of our world at the particle level. Even its apparently accidental symmetries, baryon number $B$ and lepton number $L$, are seemingly of high quality and have never been observed to be violated. One \emph{could}, however, argue that the established observation of non-zero neutrino masses is not only a sign for physics beyond the SM but also for possible lepton number violation by two units.  
This argument is based on an interpretation of the SM as an effective field theory (EFT) and the observation that the leading non-renormalizable operator is Weinberg's dimension-five operator $(\bar L H)^2/\Lambda$~\cite{Weinberg:1979sa}. This operator violates lepton number ($\Delta L =2$) and leads to Majorana neutrino masses of order $\langle H\rangle^2/\Lambda$ after electroweak symmetry breaking, which gives the correct neutrino mass scale for a cutoff $\Lambda\sim\unit[10^{14}]{GeV}$. Besides explaining neutrino oscillation data, an EFT scale this high has little impact on other observables and thus nicely accommodates the absence of non-SM-like signals in our experiments.

An ever-increasing number of renormalizable realizations of the Weinberg operator exist in the literature, the simplest of which arguably being the famous type-I seesaw mechanism~\cite{Minkowski:1977sc} that introduces three heavy right-handed neutrinos to the SM field content.
While the Weinberg operator \emph{explicitly} breaks lepton number, underlying renormalizable models could have a dynamical origin for $\Delta L =2$ via \emph{spontaneous} breaking of the global $U(1)_L$ symmetry. This leads to the same Weinberg operator and thus Majorana neutrino masses, but as a result of the spontaneous breaking of a continuous global symmetry also a Goldstone boson appears in the spectrum. This pseudo-scalar Goldstone boson of the lepton number symmetry was proposed a long time ago and was dubbed the Majoron~\cite{Chikashige:1980ui,Schechter:1981cv}.

The Majoron is obviously intimately connected and coupled to Majorana neutrinos, but at loop level also receives couplings to the other SM particles. This makes it a simple renormalizable example of an axion-like particle (ALP), defined essentially as a light pseudo-scalar with an approximate shift symmetry.  Although not our focus here, by coupling the Majoron to quarks it is even possible to identify it with the QCD axion~\cite{Mohapatra:1982tc,Langacker:1986rj,Shin:1987xc,Ballesteros:2016euj,Ballesteros:2016xej}, thus solving the strong CP problem dynamically.
The main appeal of the Majoron ALP is that its couplings are not free but rather specified by the seesaw parameters, which opens up the possibility to reconstruct the seesaw Lagrangian by measuring the Majoron couplings~\cite{Garcia-Cely:2017oco}.
This is aided by the fact that the loop-induced effective operators that couple the Majoron to the SM are only suppressed by one power of the lepton-number breaking scale $\Lambda$, whereas right-handed neutrino-induced operators without Majorons are necessarily suppressed by $1/\Lambda^2$~\cite{Broncano:2002rw,Broncano:2003fq,Broncano:2004tz,Cirigliano:2005ck,Abada:2007ux,Gavela:2009cd,Coy:2018bxr}, rendering it difficult to reconstruct the seesaw parameters in that way.

In this article we complete the program that was started in the inaugural Majoron article~\cite{Chikashige:1980ui} and derive all Majoron couplings to SM particles. The tree-level and one-loop couplings were obtained a long time ago; here we go to the two-loop level in order to calculate the remaining couplings, which include the phenomenologically important couplings to photons as well as to quarks of different generations.
Armed with this complete set of couplings we then discuss various phenomenological consequences and constraints on the parameters. This includes a discussion of Majorons as dark matter (DM).

The rest of this article is structured as follows: in Sec.~\ref{sec:tree_and_one_loop} we introduce the singlet Majoron model and reproduce the known tree-level and one-loop couplings. In Sec.~\ref{sec:two_loop} we present results of our novel two-loop calculations necessary for the Majoron couplings to gauge bosons and to quarks of different generations. The phenomenological aspects of all these couplings are discussed in Sec.~\ref{sec:pheno}. Finally, we conclude in Sec.~\ref{sec:conclusion}.

\section{Majoron couplings at tree level and one loop}
\label{sec:tree_and_one_loop}

In this article we consider the minimal \emph{singlet} Majoron model~\cite{Chikashige:1980ui},
\begin{align}
\L = -\overline{L} y N_R H -\tfrac12\overline{N}_R^c \lambda  N_R \sigma +\hc  - \text{V}(H,\sigma),
\label{eq:lagrangian}
\end{align}
which introduces three right-handed neutrinos $N_R$ coupled to the SM lepton doublets $L$ and Higgs doublet $H$, and one SM singlet complex scalar $\sigma$ carrying lepton number $L = -2$, minimally coupled to the right-handed neutrinos proportional to the Yukawa matrices $y$ and $\lambda$.  We do not specify the scalar potential $\text{V}(H,\sigma)$ but simply assume that $\sigma = (f+\sigma^0 + \i J)/\sqrt{2}$ obtains a vacuum expectation value $f$, which then gives rise to the right-handed Majorana mass matrix $M_R =  f \lambda/\sqrt2$.
$J$ is the Majoron, $\sigma^0$ is a massive CP-even scalar with mass around $f$, assumed to be inaccessibly heavy in the following. Both $M_R$ and the charged-lepton mass matrix are chosen to be diagonal without loss of generality, effectively shifting all mixing parameters into $y$. Electroweak symmetry breaking via $\langle H\rangle = (v/\sqrt2,0)^T$ yields the Dirac mass matrix $M_D = y v/\sqrt2$. The full $6\times 6$ neutrino mass matrix in the basis $(\nu_L^c, N_R) = V n_R$ is then
\begin{align}
\begin{split}
\L &= -\frac12 \bar{n}_R^c V^T \matrixx{0 & M_D \\ M_D^T & M_R} V n_R +\hc \\
&\equiv-\frac12 \bar{n}_R^c M_n n_R +\hc \,,
\end{split}
\label{eq:neutrino_mass_matrix}
\end{align}
where $V$ is the unitary $6\times 6$ mixing matrix to the states $n_R$, which form the Majorana mass eigenstates $n = n_R + n_R^c$. The diagonal mass matrix $M_n = \diag (m_1,\dots, m_6)$ consists of the physical neutrino masses arranged in ascending order.  Throughout this article, we denote mass matrices with capital letters $M_x$ and individual mass eigenvalues with small letters $m_i$.  In the mass eigenstate basis, the tree-level neutrino couplings to $J$, $Z$, $W^-$, and $h$, take the form~\cite{Pilaftsis:1993af,Pilaftsis:2008qt}
\begin{align}
\begin{split}
\L_J &= -\frac{\i J }{2f}\sum_{i,j=1}^6 \overline{n}_i \left[C_{ij} (m_i P_L - m_j P_R) \right.\\
&\qquad\qquad\qquad \left.+ C_{ji}(m_j P_L - m_i P_R) +  \delta_{ij} \gamma_5 m_i\right] n_j \,,\nonumber
\end{split}\\
\L_Z &= \frac{g_w}{4 \cos\theta_W} \sum_{i,j=1}^6\overline{n}_i \slashed{Z}\left[ C_{ij} P_L - C_{ji} P_R\right] n_j\,,  \label{eq:Jnunu}\\
\L_W &= \frac{g_w}{\sqrt2} \sum_{j=1}^6\sum_{\alpha=1}^3 \left( \overline{\ell}_\alpha  B_{\alpha j}\slashed{W}^- P_L n_j + \overline{n}_jB_{\alpha j}^* \slashed{W}^+ P_L \ell_\alpha\right) ,\nonumber\\
\begin{split}
\L_h &= -\frac{h }{2 v}\sum_{i,j=1}^6 \overline{n}_i \left[ C_{ij} (m_i P_L + m_j P_R) \right.\\
&\qquad\qquad\qquad \left.+ C_{ji}(m_j P_L + m_i P_R) \right] n_j \,,\nonumber
\end{split}
\end{align}
where $g_w = e/\sin\theta_W$ with Weinberg angle $\theta_W$ and
\begin{align}
C_{ij} \equiv \sum_{k=1}^3 V_{ki} V^*_{kj}\,, &&
B_{\alpha j} &\equiv \sum_{k=1}^{3} V_{\alpha k}^\ell V_{k j}^*=  V_{\alpha j}^* \,.
\end{align}
In the last equation we used $V^\ell_{\alpha k} =\id_{\alpha k}$ since we work in the basis where the charged-lepton mass matrix is diagonal.

The $6\times 6$ matrix $C$ and the $3\times 6$ matrix $B$ satisfy a number of identities~\cite{Pilaftsis:1991ug,Pilaftsis:1992st} that are particularly important in order to establish ultraviolet (UV) finiteness of amplitudes involving neutrino loops:
\begin{gather}
\nonumber C = C^\dagger= C C\,,\\
\begin{aligned}
B B^\dagger &= \id\,,& C M_n C^T &= 0\,,\\
B^\dagger B &= C\,,& B M_n C^T &= 0\,,\\
B C &= B\,, & B M_n B^T &= 0\,.\\
\end{aligned}\label{eq:BandCidentities}
\end{gather}

So far we have not made any assumption about the scale of $M_R$. In the following we will work in the seesaw limit $M_D \ll M_R$, resulting in a split neutrino spectrum with three heavy neutrinos with mass matrix $M_R$ and three light neutrinos with seesaw mass matrix $M_\nu \simeq -M_D M_R^{-1} M_D^T$, naturally suppressed compared to the electroweak scale $v$.  This hierarchy permits an expansion of all relevant matrices in terms of the small $3\times 3$ matrix $A\equiv U^\dagger M_D M_R^{-1}$, where $U$ is the unitary $3\times 3$ Pontecorvo--Maki--Nakagawa--Sakata (PMNS) matrix. Parametrically this corresponds to an expansion in the scale hierarchy $v\ll f$ which we refer to as the seesaw expansion. To leading order, the matrices take the form
\begin{align}
\begin{split}
C &\simeq \matrixx{ \id - A A^\dagger & A\\ A^\dagger & A^\dagger A},\\
B &\simeq \matrixx{ U(\id - \tfrac12 A A^\dagger ) & U A},\\
M_n &\simeq \matrixx{ -A M_R A^T & 0 \\ 0 & M_R} .
\end{split}
\label{eq:seesaw_expansion}
\end{align}
Note that $A M_R A^T$ is \emph{diagonal}, which imposes constraints on $A$ and provides an implicit definition of $U$. These constraints may be automatically satisfied using the Casas--Ibarra parametrization~\cite{Casas:2001sr}; however, more useful for our purpose is the Davidson--Ibarra parametrization~\cite{Davidson:2001zk}, which uses $M_\nu = -A M_R A^T$ and $M_D M_D^\dagger$ as the independent  matrices containing all seesaw parameters. Since $M_\nu$ is essentially already fixed by neutrino oscillation experiments (modulo the phases, hierarchy, and overall mass scale), the next step is to experimentally determine $M_D M_D^\dagger$. As we will see, this could in principle be achieved by measuring \emph{Majoron couplings} without ever observing the heavy right-handed neutrinos.

To this effect let us point out some interesting properties of the hermitian matrix $M_D M_D^\dagger$~\cite{Garcia-Cely:2017oco}: its determinant is simply $\det M_D M_D^\dagger =\det M_n = \prod_{j=1}^6 m_j$, which is strictly positive in the model at hand even if one of the light neutrinos were massless at tree level~\cite{Davidson:2006tg}. Thus, $M_D M_D^\dagger$ is positive definite, which yields a chain of inequalities for the off-diagonal entries $(M_D M_D^\dagger)_{ij}$, $i\neq j$ (see e.g.~Ref.~\cite{Matrix_Analysis}):
\begin{align}
\begin{split}
|(M_D M_D^\dagger)_{ij}| & < \sqrt{(M_D M_D^\dagger)_{ii}(M_D M_D^\dagger)_{jj}}\\
& \leq \frac{(M_D M_D^\dagger)_{ii}+(M_D M_D^\dagger)_{jj}}{2}\\
& \leq \frac{1}{2} \, \tr (M_D M_D^\dagger) \,.
\end{split}
\label{eq:inequality}
\end{align}
This provides a useful way to constrain magnitudes of the elements of $M_D M_D^\dagger$ since its trace appears in many couplings of the Majoron.

From Eq.~\eqref{eq:Jnunu} all loop-induced Majoron couplings are necessarily proportional to $1/f$.  But many couplings contain additional powers of $M_R^{-1}\propto 1/f$, which make them higher order in the seesaw expansion. We will neglect these suppressed couplings and focus on those that are down by only one power of $1/f$.
For the sake of generality, we determine the Majoron couplings assuming an explicit shift-symmetry-breaking Majoron mass term $-\frac{1}{2}m_J^2 J^2$, making $J$ a \emph{pseudo}-Goldstone boson. This mass could be explicit~\cite{Gu:2010ys,Frigerio:2011in} or arise from quantum-gravity effects~\cite{Akhmedov:1992hi,Rothstein:1992rh,Alonso:2017avz}. 

\subsection{Neutrino couplings}

By inserting Eq.~\eqref{eq:seesaw_expansion} into Eq.~\eqref{eq:Jnunu}, the tree-level Majoron coupling to the light active Majorana neutrinos in the seesaw limit is
\begin{align}
\L_J &= \frac{\i J }{2f}\sum_{j=1}^3 m_j \overline{n}_j\gamma_5 n_j\,.
\end{align}
These diagonal Majoron couplings to neutrinos are formally second order in the seesaw expansion since $m_{1,2,3}/f \sim M_D^2/(M_R f)\sim (v/f)^2$. The omitted off-diagonal $Jn_i n_j$ couplings are determined by the matrix $A A^\dagger A M_R A^T/f \sim (v/f)^3$, which are further suppressed, and lead to irrelevantly slow active-neutrino decays $n_i \to n_j J$ ~\cite{Schechter:1981cv}.

Assuming for simplicity $m_J \gg m_{1,2,3}$, the Majoron's partial decay rate into light neutrinos is
\begin{align}
\Gamma (J\to \nu\nu) = \frac{m_J}{16\pi f^2}\sum_{j=1}^3 m_j^2 \,. 
\label{eq:J_to_nunu}
\end{align}
For sufficiently large $f$ the Majoron becomes a long-lived DM candidate~\cite{Rothstein:1992rh,Berezinsky:1993fm,Lattanzi:2007ux,Bazzocchi:2008fh,Frigerio:2011in,Lattanzi:2013uza,Queiroz:2014yna, Wang:2016vfj}, discussed in Sec.~\ref{sec:Majoron_dark_matter}.  

As mentioned earlier, the Majoron couplings to all other SM particles are leading order in the seesaw expansion, i.e.~proportional to $1/f$, and may easily dominate the phenomenology despite the additional loop suppression~\cite{Garcia-Cely:2017oco}.  Therefore, a thorough discussion of the Majoron requires knowledge of all loop-induced couplings that are leading order in the seesaw expansion.  Using the tree-level couplings of Eq.~\eqref{eq:Jnunu} we calculate the loop-induced Majoron couplings to the rest of the SM particles and provide them below.

\subsection{Charged fermion couplings}

\begin{figure}[b]
\includegraphics[width=0.75\columnwidth]{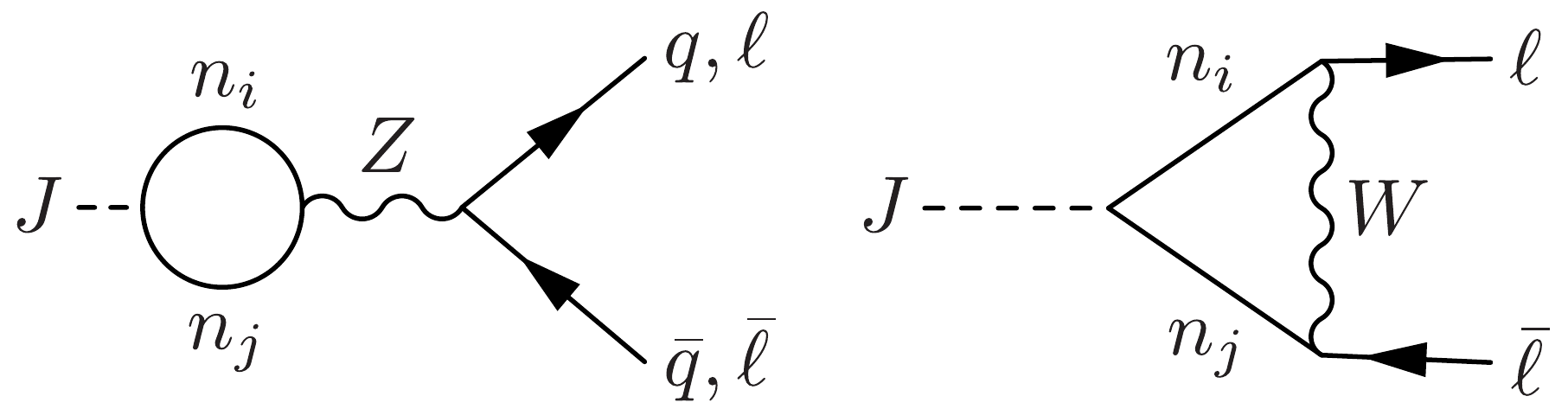}
\caption{
Loop-induced Majoron couplings to charged fermions with the Majorana neutrino mass eigenstates $n_i$ running in the loops.
}
\label{fig:Majoron_fermion_coupling}
\end{figure}

The leading order couplings to charged fermions are obtained from the one-loop diagrams in Fig.~\ref{fig:Majoron_fermion_coupling}.  These were calculated long ago, both in the one-generation case~\cite{Chikashige:1980ui} and in the three-generation case, which leads to off-diagonal Majoron couplings to leptons~\cite{Pilaftsis:1993af}.
At leading order in the seesaw expansion, these couplings take a simple form~\cite{Garcia-Cely:2017oco}, with (diagonal) quark couplings
\begin{align}
\hspace{-1ex}\L_{Jqq} = \frac{ \i J }{16\pi^2 v^2 f} \tr (M_D M_D^\dagger) \left( \bar{d} M_d \gamma_5 d-\bar{u} M_u \gamma_5 u \right),
\label{eq:Jqq}
\end{align}
and charged lepton couplings
\begin{align}
\begin{split}
\L_{J\ell\ell} &= \frac{ \i J }{16\pi^2 v^2 f} \bar{\ell} \left(M_\ell \, \tr (M_D M_D^\dagger) \gamma_5 \right.\\
&\left.\quad +2 M_\ell M_D M_D^\dagger P_L - 2 M_D M_D^\dagger M_\ell P_R \right)\ell \,,
\end{split}
\label{eq:Jellell}
\end{align}
where $M_{\ell,u,d}$ denote the diagonal mass matrices of the appropriate SM fermions.
In addition to exhibiting decoupling in the seesaw limit $M_R\sim f \to\infty$, these couplings vanish in the electroweak symmetric limit $v\to 0$ as expected since $J$ is an electroweak singlet.
The quark couplings can be used to derive the Majoron couplings to nucleons $N=(p,n)^T$, using the values from Ref.~\cite{diCortona:2015ldu}:
\begin{align}
\hspace{-1ex}\L_{JNN} \simeq \frac{ \i J \,\tr (M_D M_D^\dagger)}{16\pi^2 v^2 f} \bar{N}\matrixx{-1.30 m_p & 0\\0 & 1.24 m_n} \gamma_5 N \,.
\label{eq:Jpp}
\end{align}

At this point let us make some remarks about CP violation. Already in the one-loop processes above one encounters loop-induced Majoron mixing with the Brout--Englert--Higgs boson $h$, which would result in Majoron couplings to the scalar bilinear $\bar{f}f$ as opposed to the pseudo-scalar $\bar{f}\i \gamma_5 f$. It was noted in Ref.~\cite{Pilaftsis:1993af} that the relevant $J$--$h$ mixing diagrams vanish for $m_J =0$. For $m_J\neq 0$ the $J$--$h$ amplitude is of order $(v/f)^2$ in seesaw and hence negligible. This can be understood by noting that CP-violating phases in the Davidson--Ibarra parametrization reside both in the active-neutrino mass matrix $M_\nu = -A M_R A^T$ and in the off-diagonal entries of the hermitian matrix $M_D M_D^\dagger$, each containing three complex phases~\cite{Davidson:2001zk}. CP-violating Majoron couplings via $M_\nu$ are unavoidably suppressed by $M_\nu/f\sim (v/f)^2$, leaving only  $M_D M_D^\dagger$ as a potential source. However, closing lepton loops implies an amplitude dependence on  $\tr (M_D M_D^\dagger g(M_\ell))$, with some function $g(M_\ell)$ of the charged-lepton mass matrix. Since the latter is diagonal, the trace depends only on the real diagonal entries of $M_D M_D^\dagger$, resulting in an effectively CP-conserving amplitude.
The CP phases of $M_D M_D^\dagger$ thus only appear in the off-diagonal Majoron couplings to leptons, at least to lowest order in the seesaw expansion.

\subsection{Couplings to gauge bosons}

\begin{figure}[t]
\includegraphics[width=\columnwidth]{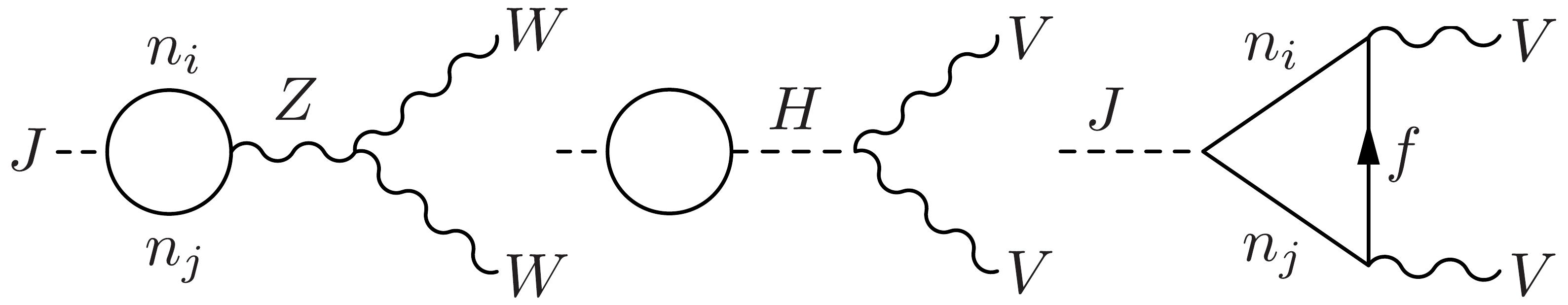}
\caption{
Loop-induced Majoron couplings to $WW$ and $ZZ$ at one loop.  These diagrams generate couplings that are subdominant in the seesaw expansion.}
\label{fig:JVV_1loop}
\end{figure}

At one-loop order, the only non-vanishing couplings to gauge bosons are to $WW$ and $ZZ$, with typical diagrams shown in Fig.~\ref{fig:JVV_1loop}.  However, they are higher order in the seesaw expansion, which can be understood as follows: the amplitudes come with a factor of $M_D M_D^\dagger/f$ in order to achieve the necessary $N_R$--$\nu_L$ mixing to close the loop; on dimensional grounds there is an additional $M_R^{-2}$ suppression since this is the only high mass scale in the loop.  Explicit one-loop formulae can be found in Refs.~\cite{Latosinski:2012qj,Latosinski:2012ha}.  The leading seesaw behavior for coupling to gauge bosons without the $M_R^{-2}$ suppression starts at two-loop order.

At this point it is appropriate to discuss the connection to \emph{anomalies} in the minimal Majoron model. There is some confusion in the literature regarding the question of whether the Majoron is the Goldstone boson of the anomaly-free $U(1)_{B-L}$ or the anomalous $U(1)_L$. Both choices seem equally valid because baryon number remains unaltered by the Lagrangian in Eq.~\eqref{eq:lagrangian}. Since according to common lore Goldstone couplings to gauge bosons are determined by the anomaly structure of the theory, this leads to a paradox when attempting to guess the form of Majoron couplings to $W$ and $Z$. 
The resolution was recently presented in Ref.~\cite{Quevillon:2019zrd}, where it was explained that Goldstone couplings to gauge bosons are driven entirely by \emph{non-anomalous} processes. Anomalies still serve as a useful bookkeeping device for the couplings to vector-like gauge bosons such as gluons and photons, but fail for chiral gauge bosons.
Disregarding anomalies it is then necessary to calculate Goldstone couplings to gauge bosons in perturbation theory, the results of which we present in the next section.
Additionally, we emphasize that the non-vanishing Majoron coupling to electroweak gauge bosons (Eqs.~\eqref{eq:JZgammaeffCoup}, \eqref{eq:JZZeffCoup}, and \eqref{eq:JWWeffCoup} below) does not lead to nonperturbative violation of the shift symmetry beyond $m_J$ due to the absence of electroweak instantons without $B+L$ violation \cite{Anselm:1992yz,Anselm:1993uj,Perez:2014fja}. Therefore, electroweak instantons cannot generate Majoron mass. 

\section{Majoron couplings at two loops}
\label{sec:two_loop}

In this section we present Majoron couplings for which the leading seesaw behavior arises at two loops. To automate the evaluation of some $\mathcal{O}(100)$ Feynman diagrams contributing to each effective coupling we implemented this model in Feynman gauge including all Goldstone bosons~\cite{Pilaftsis:2008qt} in \texttt{FeynRules}~\cite{Alloul:2013bka}, and generated the necessary amplitudes with \texttt{FeynArts}~\cite{Hahn:2000kx}. We validated our implementation by reproducing the tree-level and one-loop couplings above.

The Feynman diagrams naturally divide into two sets (see Fig.~\ref{fig:JVVdiagrams}).  Set I diagrams contain the one-particle irreducible (1PI) two-loop diagrams.  Set II diagrams contain the reducible diagrams that are dominated by $J$--$Z$ mixing.  We used an in-house \emph{Mathematica} implementation of \emph{expansion by regions} as described in Ref.~\cite{Smirnov:1999bza} to carry out a double asymptotic expansion $m_{4,5,6}\rightarrow\infty$, $m_{1,2,3}\rightarrow0$ of the two-loop vertex integrals, and to algebraically reduce the one-loop \cite{Passarino:1978jh} and two-loop \cite{Davydychev:1995nq} tensor integrals in dimensional regularization.  We treated $\gamma_5$ naively, such that it anti-commutes with all other Dirac matrices while also preserving the cyclic property of traces.  Finally, after expanding around four spacetime dimensions, we summed over fermion generations to extract the leading seesaw behavior of each Majoron coupling.  We found the couplings to be expressible as simple sums of one-loop functions and rational terms.  In principle, two-loop self-energy and vacuum integrals may be present, but they cancel away in the course of reduction, leaving behind rational terms.

We have checked our results by confirming that all amplitudes are proportional to the expected tensor structures, are UV finite upon using the relations of Eq.~\eqref{eq:BandCidentities}, and have the expected limiting low-energy/small-mass behavior.  Additionally, we confirmed that our results are insensitive to the treatment of $\gamma_5$ by reevaluating them in several different ways, including projecting the integrals onto form factors and also starting from cyclically reordered Dirac traces, and finding the same answer upon expanding around four spacetime dimensions.  

We pause to comment on how we quote our results for couplings to gauge bosons $\{VV'\} = \{gg, \gamma\gamma, Z\gamma, ZZ, W^+W^-\}$.  We phrase our results in terms of on-shell decay amplitudes
\begin{multline}
\mathcal{M}(J\rightarrow V(k_1) V'(k_2)) = \\
- g_{JVV'} \epsilon^{\mu\nu\rho\sigma} \epsilon^*_{\mu}(k_1)\epsilon^*_{\nu}(k_2) k_{1,\rho} k_{2,\sigma} \,.
\end{multline}
It is commonplace to see these amplitudes interpreted as effective couplings as they \emph{appear} to match onto EFT operators of the form~\cite{Brivio:2017ije}
\begin{equation}\label{eq:naiveJVVoperator}
\mathcal{L} = -\frac{g_{JVV'}}{4}J V^{\mu\nu}\tilde{V}'_{\mu\nu}\,,
\end{equation}
where $V^{\mu\nu}$ is the appropriate field-strength tensor and $\tilde{V}^{\mu\nu}$ its dual.  However, we caution the reader that the identification with effective couplings in this way is somewhat clumsy for the following reasons.  First, matching onto local operators should be carried out for off-shell Green functions which have been expanded in the external momenta.  Second, our effective couplings $g_{JVV'}$ cannot be viewed in a Wilsonian sense, since degrees of freedom \emph{lighter} than the Majoron contribute in certain mass ranges, nor can it be viewed in the 1PI sense since the couplings include Set II diagrams that are not one-particle irreducible.  Therefore, interpreting our results as coefficients of effective operators should be done with care.

\subsection{Coupling to two gluons}
\begin{figure}[t]
\includegraphics[width=0.80\columnwidth]{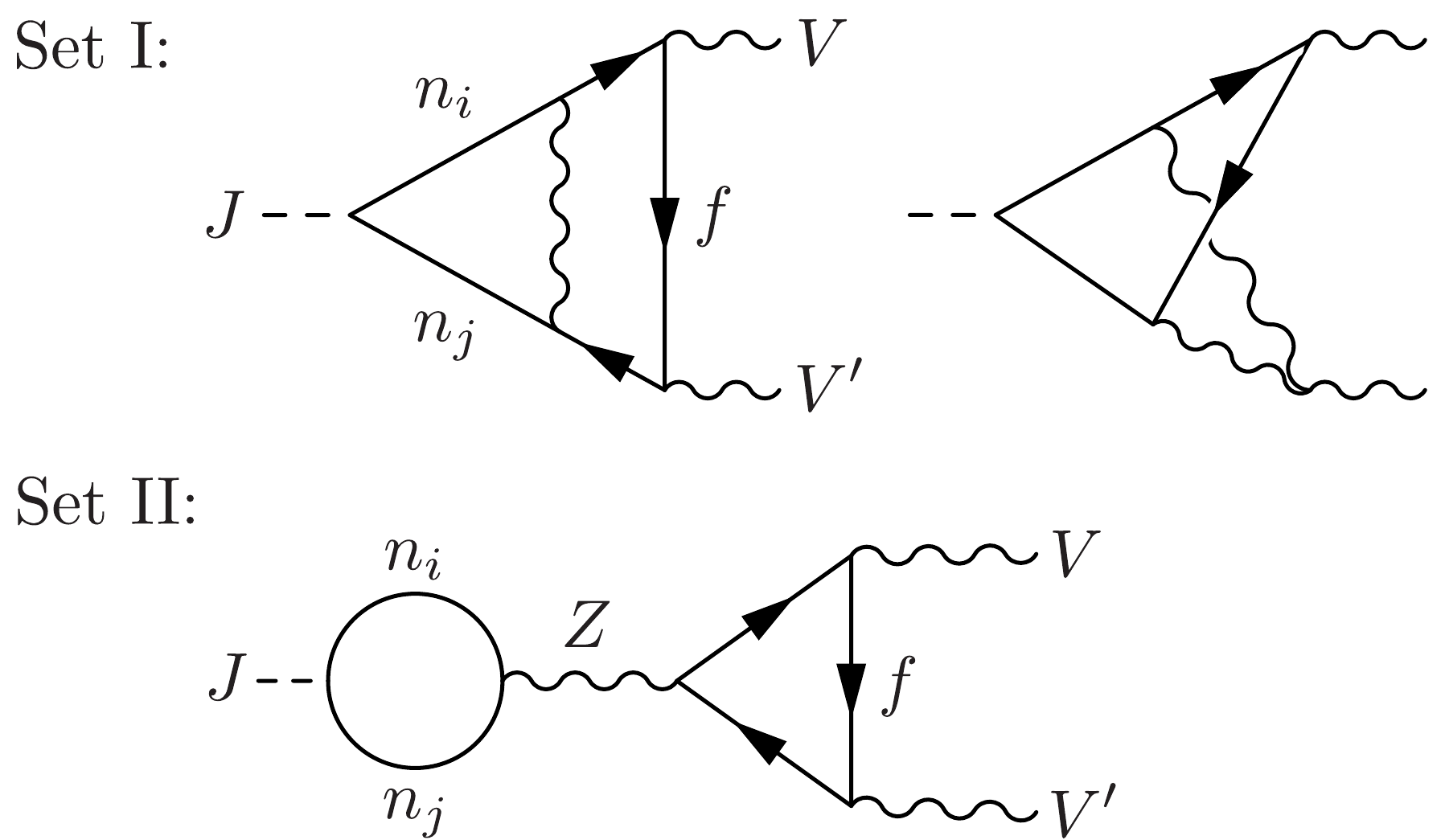}
\caption{
Representative two-loop diagrams contributing to loop-induced Majoron couplings to vector bosons $\{VV'\} = \{gg, \gamma\gamma, Z\gamma, ZZ, W^+W^-\}$.  Set I contains the two-loop 1PI diagrams and Set II contains reducible diagrams dominated by $J$--$Z$ mixing.  Here, $n_i$ and $n_j$ are Majorana neutrino mass eigenstates and $f$ SM fermions (not necessarily all identical).}
\label{fig:JVVdiagrams}
\end{figure}

Assuming a sufficiently heavy Majoron $m_J \gtrsim \Lambda_\text{QCD}$, the coupling to free gluons comes entirely from $J$--$Z$ mixing diagrams of Set II in Fig.~\ref{fig:JVVdiagrams}.  A straightforward evaluation of the decay amplitude $J\to gg$ at leading order in the seesaw expansion yields the simple expression
 \begin{align}
 g_{J g g} = \frac{\alpha_S}{16 \pi ^3 v^2 f} \tr(M_D M_D^\dagger)\sum_{q=u,d} T_3^q \  h\Big(\frac{m_J^2}{4 m_q^2}\Big)\,,
 \end{align}
 with $T_3^u = -T_3^d = 1/2$ and the loop function
 \begin{align}
\begin{split}
h(x) &\equiv -\frac{1}{4 x} \big(\log [1-2 x + 2\sqrt{x (x-1)}]\big)^2 -1\\
& =
\begin{cases}
 \frac{x}{3} + \frac{8x^2}{45} + \frac{4 x^3}{35}+\mathcal{O}(x^4)\,, & x\to0\,,\\
 -1 + \frac{(\pi +\i \log (4 x))^2}{4 x}+\mathcal{O}(x^{-2})\,, & x\to\infty\,.
\end{cases}
\end{split}
\end{align}
For small $m_J$, the amplitude vanishes as $g_{J g g} \sim m_J^2$ and indicates that at leading order in the derivative expansion the amplitude matches onto an operator $(\partial^2J) G^{a \mu\nu}\tilde{G}^a_{\mu\nu}$ instead of $J G^{a \mu\nu} \tilde{G}^{a}_{\mu\nu}$ as in Eq.~\eqref{eq:naiveJVVoperator}. This implies that the Majoron does not solve the strong CP problem, as this operator is insensitive to a constant shift $J\to J + c$ that could otherwise be used to cancel the strong CP $\theta$ term~\cite{Quevillon:2019zrd}.  Furthermore, contrary to the claim in Ref.~\cite{Latosinski:2012qj}, Majorons without tree-level couplings to quarks cannot solve the strong CP problem even at higher loop order~\cite{Latosinski:2015pba,Nakawaki:2018cwk}.

\subsection{Coupling to two photons}
\label{sec:diphoton}

The Majoron coupling to photons at two-loop level receives contributions from both sets of Feynman diagrams in Fig.~\ref{fig:JVVdiagrams},
\begin{equation}
g_{J\gamma\gamma} = g_{J\gamma\gamma}^\text{I} + g_{J\gamma\gamma}^\text{II}\,,
\end{equation}
and yields the partial decay rate into two photons,
\begin{align}
\Gamma (J\to\gamma\gamma) = \frac{|g_{J\gamma\gamma}|^2 m_J^3}{64\pi} \,.
\end{align}

The contributions from Set II were calculated in Ref.~\cite{Garcia-Cely:2017oco} with the result
  \begin{align}
\hspace{-1.7ex} g_{J\gamma\gamma}^\text{II} = \frac{\alpha}{8 \pi ^3 v^2 f} \tr(M_D M_D^\dagger) \sum_f  N_c^f Q_f^2 T_3^f \ h\Big(\frac{m_J^2}{4 m_f^2}\Big) \,,
\label{eq:JgammagammaSet1}
 \end{align}
already simplified with the help of the electroweak anomaly cancellation condition $\sum_f  N_c^f Q_f^2 T_3^f=0$. Here, $ N_c^{u,d} = 3=  3 N_c^\ell$ is the number of colors, $T_3^{u}=1/2 = - T_3^{d,\ell}$ the isospin, and $\{Q_\ell, Q_d, Q_u\}=\{-1,-1/3,+2/3\}$ the electric charge in units of $e=\sqrt{4\pi \alpha}$. We complete the evaluation of $g_{J\gamma\gamma}$ here by computing the additional contributions arising from Set~I diagrams, which give
 \begin{align}
g_{J\gamma\gamma}^\text{I} = \frac{\alpha}{8 \pi ^3 v^2 f}   \sum_\ell (M_D M_D^\dagger)_{\ell\ell}  \ h\Big(\frac{m_J^2}{4 m_\ell^2}\Big)  \,.
 \end{align}
Just as for the gluon coupling, the amplitude vanishes as $g_{J\gamma\gamma}\sim m_J^2$  for small Majoron masses, implying that the leading effective operator this amplitude matches onto in the derivative expansion is $(\partial^2 J) F_{\mu\nu} \tilde{F}^{\mu\nu}$ rather than the typically occurring $J F_{\mu\nu} \tilde{F}^{\mu\nu}$.

\begin{figure*}[t]
	\includegraphics[width=0.88\textwidth]{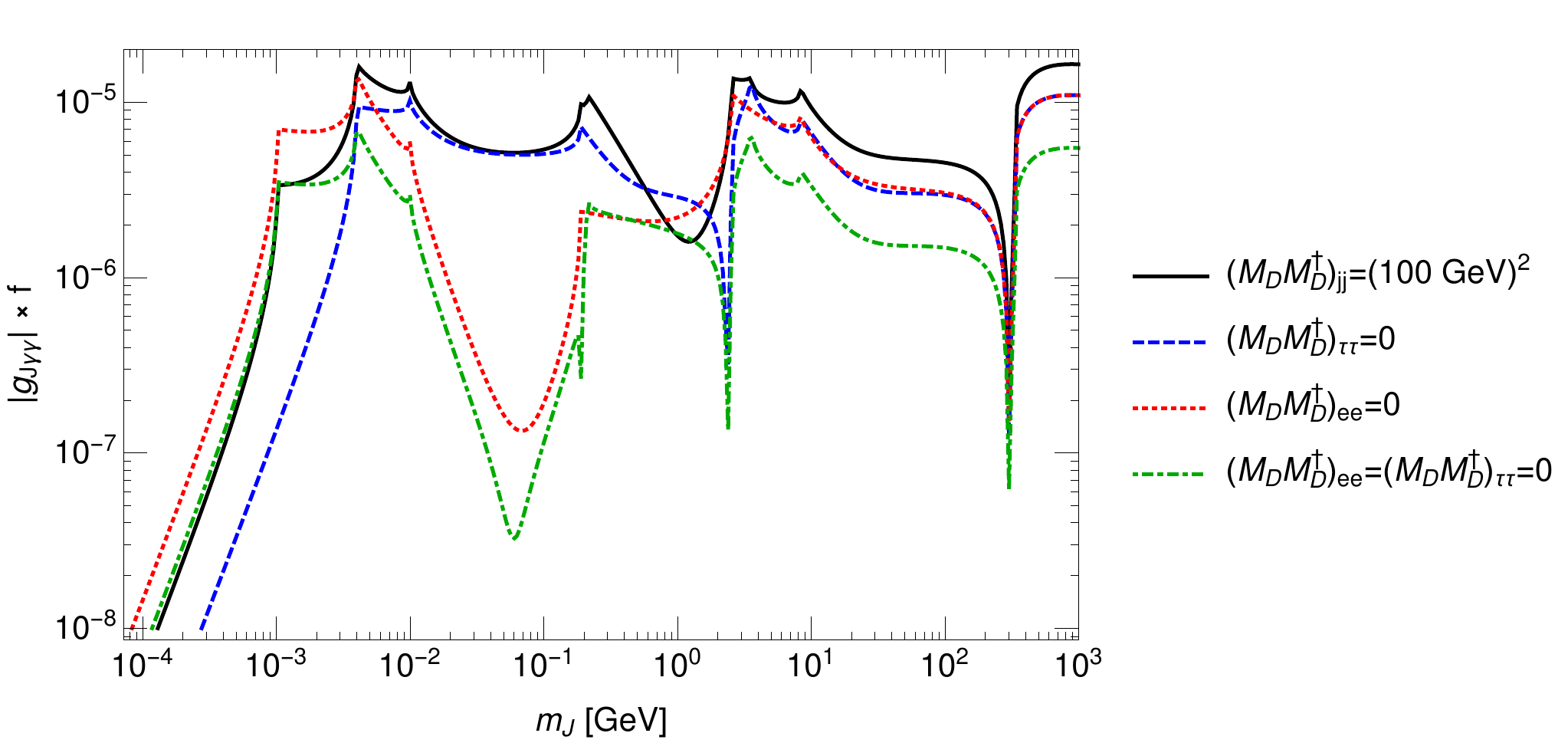}
	\caption{
		$J\gamma\gamma$ coupling for various hierarchies in the diagonal elements of $M_D M_D^\dagger$. The black solid line shows $(M_D M_D^\dagger)_{ee}=(M_D M_D^\dagger)_{\mu\mu}=(M_D M_D^\dagger)_{\tau\tau}=(\unit[100]{GeV})^2$; the blue dashed line shows $(M_D M_D^\dagger)_{ee}=(M_D M_D^\dagger)_{\mu\mu}=(\unit[100]{GeV})^2$, $(M_D M_D^\dagger)_{\tau\tau}=0$ (which corresponds to $g_{Jee}=g_{J\mu\mu}=0$); the red dotted line shows $(M_D M_D^\dagger)_{\mu\mu}=(M_D M_D^\dagger)_{\tau\tau} = (\unit[100]{GeV})^2$, $(M_D M_D^\dagger)_{ee}=0$; and the green dot-dashed line shows $(M_D M_D^\dagger)_{\mu\mu}= (\unit[100]{GeV})^2\gg (M_D M_D^\dagger)_{ee}$, $(M_D M_D^\dagger)_{\tau\tau}$.
	}
	\label{fig:Jgammagamma_amplitude}
\end{figure*}

For $m_J \ll m_f$ we can relate the total Majoron--photon coupling $g_{J\gamma\gamma}$ to the dimensionless diagonal fermion couplings of Eqs.~\eqref{eq:Jqq} and~\eqref{eq:Jellell}, $g_{Jff}\,J\bar{f}\i\gamma_5 f$, as
   \begin{align}
g_{J\gamma\gamma}&\simeq -\frac{\alpha m_J^2}{12 \pi } \sum_f N_c^f Q_f^2 \frac{ g_{Jff}}{m_f^3} \,,
 \end{align}
which agrees with the EFT result of Ref.~\cite{Nakayama:2014cza}.  Since the $g_{J f f}$ couplings can have different signs and magnitudes, the $J\gamma\gamma$ coupling for $m_J<m_e$ could be heavily suppressed.  The key point and crucial result of this full two-loop calculation is that the $J\gamma\gamma$ coupling has a richer structure than anticipated in Ref.~\cite{Garcia-Cely:2017oco} based on the evaluation of $g_{J\gamma\gamma}^\text{II}$ alone.  This is illustrated in Fig.~\ref{fig:Jgammagamma_amplitude} where we show $|g_{J\gamma\gamma}|\times f$ for a variety of hierarchies of the diagonal entries $(M_D M_D^\dagger)_{\ell\ell}$. The SM-fermion mass thresholds together with the different signs in $g_{Jff}$ potentially suppress $g_{J\gamma\gamma}$ by orders of magnitude. The typical size of the coupling for $m_J>\unit{MeV}$ is $|g_{J\gamma\gamma}|\sim 10^{-5} f^{-1} (M_D M_D^\dagger)/(\unit[100]{GeV})^2$, simply due to the unavoidable suppression factor $\alpha/(8\pi^3)$.
 In Fig.~\ref{fig:Jgammagamma_amplitude} we used the current-quark masses to evaluate $g_{J\gamma\gamma}$; for $m_J \lesssim \Lambda_\text{QCD}$ they should be replaced by hadronic loops. We have not attempted this, but we refer the interested reader to standard axion literature on the topic~\cite{Georgi:1986df,Bauer:2017ris,Alonso-Alvarez:2018irt}.

\begin{widetext}
\subsection{Coupling to \texorpdfstring{$Z$}{Z} and photon}

Next we present the $Z$-photon coupling $g_{J Z\gamma} = g_{J Z\gamma}^\text{I}+g_{J Z\gamma}^\text{II}$, which receives contributions from Set I and Set II diagrams in Fig.~\ref{fig:JVVdiagrams}. The results are
\begin{align}
g_{J Z\gamma}^\text{I} &= -\frac{\alpha}{16 \pi^3 c_W s_W v^2 f} \left(\tr(M_D M_D^\dagger) -(1-4 s_W^2) \sum_\ell (M_D M_D^\dagger)_{\ell\ell} \frac{m_J^2 h\left(\frac{m_J^2}{4 m_\ell^2}\right)-m_Z^2 h\left(\frac{m_Z^2}{4 m_\ell^2}\right)}{m_J^2-m_Z^2} \right) ,\\
g_{J Z\gamma}^\text{II} &= -\frac{\alpha}{16 \pi^3 c_W s_W v^2 f}  \tr(M_D M_D^\dagger) \sum_f 2 N_c^f Q_f T_3^f (2 Q^f s_W^2-T_3^f) \frac{m_J^2 h\left(\frac{m_J^2}{4 m_f^2}\right)-m_Z^2 h\left(\frac{m_Z^2}{4 m_f^2}\right)}{m_J^2-m_Z^2} \,,
\end{align}
with $c_W \equiv \cos\theta_W$ and $s_W \equiv \sin\theta_W$.  We have used $\sum_f  N_c^f Q_f^2 T_3^f=0=\sum_f  N_c^f Q_f (T_3^f)^2$ to simplify the formula.  
In the limit $m_J, m_Z \to 0$, the amplitude is non-vanishing,
\begin{equation}\label{eq:JZgammaeffCoup}
g_{JZ\gamma} \sim -\frac{\alpha\, \tr(M_D M_D^\dagger)}{16\pi^3 c_W s_W v^2 f}\,,
\end{equation}
and matches onto the effective operator $J Z^{\mu\nu}\tilde{F}_{\mu\nu}$, as in Eq.~\eqref{eq:naiveJVVoperator}.

\subsection{Coupling to two \texorpdfstring{$Z$}{Z} bosons}
The Majoron coupling $g_{JZZ} = g_{JZZ}^\text{I} + g_{JZZ}^\text{II}$ to two $Z$ bosons receives contributions from Sets I and II:
\begin{multline}
g_{JZZ}^\text{I} = -\frac{\alpha  }{32 \pi ^3 c_W^2 s_W^2 v^2 f}\Big[ \big(1-2 s_W^2\big)^2 \tr(M_D M_D^\dagger)
 -\frac{2}{m_J^2-4 m_Z^2} \sum_\ell (M_D M_D^\dagger)_{\ell\ell} m_\ell^2 \Big(g(\textstyle\frac{m_J^2}{4 m_\ell^2})-g(\textstyle\frac{m_Z^2}{4 m_\ell^2})\\
+\Big( 2 s_W^2(1-2 s_W^2)m_J^2 + (1-4s_W^2)^2m_Z^2\Big) C_0(m_J^2,m_Z^2,m_Z^2,m_\ell,m_\ell,m_\ell)\Big)\Big]\,,
\end{multline}
\begin{multline}
g_{JZZ}^\text{II} = \frac{\alpha}{4 \pi ^3 c_W^2 s_W^2 v^2 f}\tr(M_D M_D^\dagger) \frac{1}{m_J^2-4 m_Z^2}  \sum_f N_c^f T_3^f m_f^2 \Big[\big(T_3^f\big)^2\big(g(\textstyle\frac{m_J^2}{4 m_f^2})-g(\textstyle\frac{m_Z^2}{4 m_f^2})\big) \\
+  \Big(s_W^2  Q_f  \big(T_3^f - Q_f s_W^2\big) m_J^2
+ \big(T_3^f-2 Q_f s_W^2\big)^2 m_Z^2 \Big)C_0(m_J^2,m_Z^2,m_Z^2,m_f,m_f,m_f)
\Big]\,,
\end{multline}
where
\begin{align}
g(x) &\equiv \sqrt{\textstyle 1-\frac{1}{x}} \log[1-2x+2\sqrt{x(x-1)}]
\end{align}
and $C_0$ is the scalar three-point Passarino--Veltman function.  Despite the appearance of $(m_J^2 - 4m_Z^2)^{-1}$, the amplitude is regular at threshold $m_J\to 2 m_Z$.  The coupling $g_{JZZ}$ is nonvanishing in the limit $m_J, m_Z \to 0$,
\begin{equation}\label{eq:JZZeffCoup}
g_{JZZ} \sim -(1-3 s_W^2) \frac{\alpha\, \tr(M_D M_D^\dagger)}{48 \pi ^3 c_W^2 s_W^2 v^2 f} \,,
\end{equation}
and matches onto $J Z^{\mu\nu}\tilde{Z}_{\mu\nu}$.

\subsection{Coupling to two \texorpdfstring{$W$}{W} bosons}

Finally, we present results for the two-loop amplitude for $J\to W^+ W^-$ in order to extract the coupling $g_{JWW}$, which receives contributions from diagrams in Set I and Set II
\begin{equation}
g_{JWW} = g_{JWW}^\text{I} + g_{JWW}^\text{II,$\ell$} + g_{JWW}^\text{II,$d$} + g_{JWW}^\text{II,$u$}\,,
\end{equation}
where we have separated the Set~II $J$--$Z$ mixing contributions based on the type of SM fermions running in the loop.
The Set I diagrams give
\begin{align}
\begin{split}
g_{JWW}^\text{I} &= \frac{\alpha }{64 \pi^3 s_W^2 v^2 f}\,\frac{1}{m_J^2-4 m_W^2} \sum_\ell \frac{(M_D M_D^\dagger)_{\ell\ell}}{m_W^2}
\left\{
4 m_W^4- m_J^2 m_W^2 + 4 m_\ell^2  m_W^2 g\Big(\frac{m_J^2}{4 m_\ell^2}\Big) \right.\\
&\quad \left. + 4 m_\ell^2  (m_W^2-m_\ell^2) \left[ -\log\big(\frac{m_\ell^2}{m_\ell^2-m_W^2}\big) + m_W^2 C_0(m_J^2,m_W^2,m_W^2,m_\ell,m_\ell,0) \right]
\right\} ,
\end{split}
\end{align}
the Set II $J$--$Z$ mixing diagrams with two charged leptons in the loop give
\begin{align}
\begin{split}
g_{JWW}^\text{II,$\ell$} &= - \frac{\alpha}{32 \pi ^3 s_W^2  v^2 f} \,\frac{\tr (M_D M_D^\dagger) }{m_W^2}\frac{1}{m_J^2-4 m_W^2}\sum_\ell m_\ell^2
\left\{m_W^2 g\Big(\frac{m_J^2}{4 m_\ell^2}\Big) \right.\\
&\quad \left.+ (m_W^2-m_\ell^2) \left[ -\log\big(\frac{m_\ell^2}{m_\ell^2-m_W^2}\big) + m_W^2 C_0(m_J^2,m_W^2,m_W^2,m_\ell,m_\ell,0) \right]
\right\} ,
\end{split}
\end{align}
the Set II $J$--$Z$ mixing diagrams with two down quarks in the loop give
\begin{align}
\begin{split}
g_{JWW}^\text{II,$d$} &= - \frac{3\, \alpha }{32 \pi ^3 s_W^2  v^2 f} \frac{\tr (M_D M_D^\dagger)}{m_W^2} \frac{1}{m_J^2-4 m_W^2}\sum_{i,j} |(V_q)_{ji}|^2 m_{d_i}^2
\left\{m_W^2 \left[g\Big(\frac{m_J^2}{4 m_{d_i}^2}\Big) - g\Big(\frac{m_W^2}{m_{d_i}^2},\frac{m_W^2}{m_{u_j}^2}\Big)\right]  \right.\\
&\quad \left.+ (m_W^2-m_{d_i}^2+m_{u_j}^2) \left[ \log\big(\frac{m_{u_j}}{m_{d_i}}\big) + m_W^2 C_0(m_J^2,m_W^2,m_W^2,m_{d_i},m_{d_i},m_{u_j}) \right]
\right\} ,
\end{split}
\end{align}
and the Set II $J$--$Z$ mixing diagrams with two up quarks in the loop give
\begin{align}
\begin{split}
g_{JWW}^\text{II,$u$} &=  \frac{3\, \alpha}{32 \pi ^3 s_W^2 v^2 f} \frac{\tr (M_D M_D^\dagger)}{ m_W^2} \frac{1}{m_J^2-4 m_W^2}\sum_{i,j} |(V_q)_{ij}|^2 m_{u_i}^2
\left\{m_W^2 \left[g\Big(\frac{m_J^2}{4 m_{u_i}^2}\Big) - g\Big(\frac{m_W^2}{m_{u_i}^2},\frac{m_W^2}{m_{d_j}^2}\Big)\right] \right.\\
&\quad \left.+ (m_W^2-m_{u_i}^2+m_{d_j}^2) \left[ \log\big(\frac{m_{d_j}}{m_{u_i}}\big) + m_W^2 C_0(m_J^2,m_W^2,m_W^2,m_{u_i},m_{u_i},m_{d_j}) \right]
\right\} .
\end{split}
\end{align}
Here, $V_q$ is the unitary Cabibbo--Kobayashi--Maskawa (CKM) mixing matrix.
In the last two formulae, the two-argument loop function is
\begin{equation}
g(x,y) = \sqrt{(x+y-1)^2-4xy}\,\log\left(\frac{x+y-1}{2\sqrt{xy}}+\sqrt{\frac{(x+y-1)^2}{4xy}-1}\right) .
\end{equation}
The coupling is regular at threshold $m_J\to 2m_W$, and nonvanishing in the limit $m_J, m_W \rightarrow 0$,
\begin{equation}\label{eq:JWWeffCoup}
g_{JWW} \sim -\frac{3\alpha \,  \tr (M_D M_D^\dagger)}{128 \pi ^3 s_W^2 v^2 f}\left[ 1 - \sum_{i,j} |(V_q)_{ij}|^2\left( \frac{m_{u_i}^4-m_{d_j}^4+2 m_{d_j}^2 m_{u_i}^2 \log (m_{d_j}^2/m_{u_i}^2)}{(m_{u_i}^2-m_{d_j}^2)^2} \right) \right] ,
\end{equation}
and matches onto $JW^{+\,\mu\nu}W^-_{\mu\nu}$.

\end{widetext}

\subsection{Coupling to \texorpdfstring{$\gamma$}{photon}-Higgs and \texorpdfstring{$Z$}{Z}-Higgs}

Besides the usually considered pseudo-scalar couplings to two gauge bosons discussed above, CP-invariance also allows couplings of $J$ to $h Z$ and $h\gamma$. The former arises already at one-loop level but is seesaw suppressed; the dominant contributions to both couplings then arise at two-loop level. Because of the large number of diagrams and low phenomenological relevance of decays such as $h\to \gamma J$ compared to the processes derived above we will not, however, present the results here.

\subsection{Flavor changing quark couplings}

\begin{figure}[t]
\includegraphics[width=0.80\columnwidth]{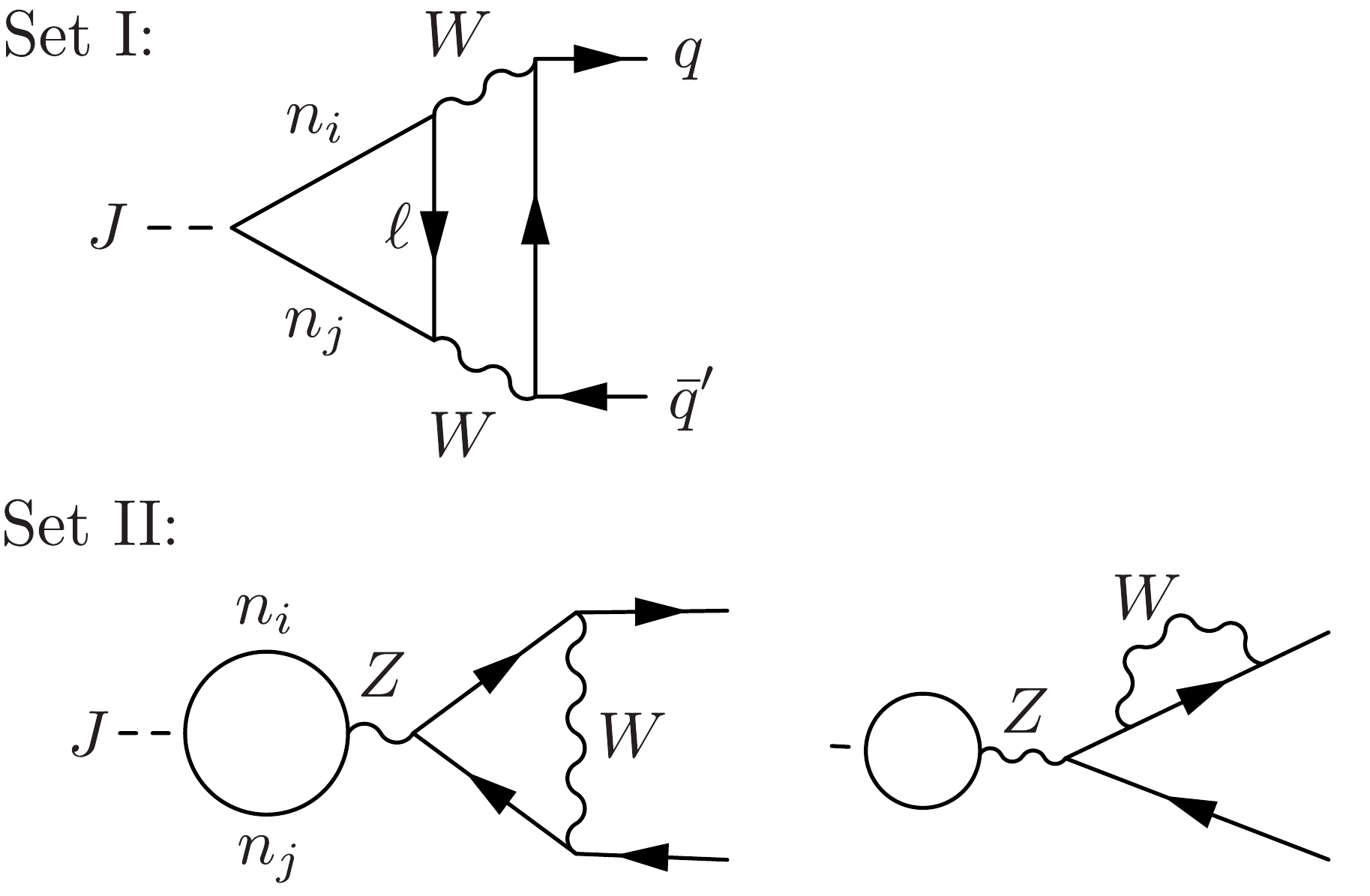}
\caption{
Representative two-loop diagrams for off-diagonal Majoron couplings to quarks.}
\label{fig:Jds}
\end{figure}

At the two-loop level we find off-diagonal Majoron couplings to quarks, which can lead, for example, to $s\to d J$ or $K\to \pi J$ at hadron level. Such flavor-changing couplings have long been advocated to search for light bosons and axions~\cite{Frere:1981cc,Hall:1981bc} and have enjoyed increased attention in recent years~\cite{Izaguirre:2016dfi,Dolan:2017osp,Dobrich:2018jyi,Gavela:2019wzg}, partly because of an improved reach at existing and upcoming experiments such as NA62 and Belle II.

In the Majoron model the relevant flavor-changing quark-level couplings arise at two loops and involve a large number of diagrams; see Fig.~\ref{fig:Jds}.
The leading logarithmic contribution to coupling to down quarks is
\begin{multline}
\mathcal{L}_{Jdd'} = -\frac{1}{128\pi^4 v^4 f}
\tr\Big(M_D \log\big(\textstyle\frac{M_R}{m_W}\big) M_D^\dagger\Big) \\
  \times (\i J\, \bar d_R  M_d V_q^\dagger M_u^2 V_q \, d_L + \hc) \,,
\end{multline}
arising from the Set I diagrams in Fig.~\ref{fig:Jds}.
The couplings are of minimal-flavor-violating (MFV) type~\cite{DAmbrosio:2002vsn}, as expected from the fact that $J$ is quark-flavor blind.
The subleading contribution from the remaining diagrams is given by
\begin{equation}
\L_{Jdd'}^\text{sub} =  \frac{- \tr (M_D  M_D^\dagger)}{512\pi^4 v^4 f}
  (\i  J\, \bar d_R  M_d V_q^\dagger F_u V_q \, d_L + \hc) \,,
\end{equation}
where the diagonal matrix $F_u$ has entries
\begin{align}
\begin{split}
F_u &= \frac{7 m_u^4 + 3 m_u^2 m_W^2 - 8 m_W^4}{m_u^2-m_W^2} +4 m_u^2 \, g\Big(\frac{m_J^2}{4m_u^2}\Big) \\
&\quad +\frac{2m_u^2 (m_u^4 -2 m_u^2 m_W^2 + 2 m_W^4)}{(m_u^2-m_W^2)^2}\, \log\Big(\frac{m_W^2}{m_u^2}\Big)\\
&\quad -4 m_u^2 m_W^2 C_0(0,0,m_J^2,m_u,m_W,m_u) \,.
\end{split}
\end{align}
For sub-TeV Majoron mass and large-enough right-handed neutrino masses, $\log\left( M_R/m_W\right) \gtrsim 1$ as is assumed in our seesaw expansion, the leading logarithmic contribution $\L_{Jdd'}$ dominates over $\L_{Jdd'}^\text{sub}$, which we neglect in the following for simplicity.

The MFV matrix $ M_d V_q^\dagger M_u^2 V_q$ relevant for $\L_{Jdd'}$ makes it clear that off-diagonal terms would vanish if all up quarks were degenerate, so these terms are necessarily proportional to up-quark mass differences. Numerically this matrix evaluates to
\begin{align}
 \hspace{-1ex}\left| \frac{M_d V_q^\dagger M_u^2 V_q}{v^3}\right| \simeq \matrixx{
0 & 0 & 8\times 10^{-8}\\ 
7\times 10^{-8} & 3\times 10^{-7} & 8\times 10^{-6}\\ 
7\times 10^{-5} & 3\times 10^{-4} & 8\times 10^{-3}
} ,
\end{align}
keeping only the largest entries.
The biggest amplitude is therefore $b\to s J$, which is, however, experimentally less clean than $s\to d J$, further discussed in Sec.~\ref{sec:light_Majorons}.

From the dominant flavor-changing down-quark couplings in $\L_{Jdd'}$ we immediately obtain the corresponding flavor-changing \emph{up-quark} couplings as
\begin{multline}
\mathcal{L}_{Juu'} = -\frac{1}{128\pi^4 v^4 f}
\tr\Big(M_D \log\big(\textstyle\frac{M_R}{m_W}\big) M_D^\dagger\Big) \\
  \times (\i J \, \bar u_R  M_u V_q M_d^2 V_q^\dagger \, u_L + \hc) \,,
\end{multline}
with the markedly smaller MFV coupling matrix
\begin{align}
 \hspace{-2.5ex}\left| \frac{M_u V_q M_d^2 V_q^\dagger}{v^3}\right| \simeq \matrixx{
0 & 0 & 0\\
3\!\times\!10^{-10} & 3\times 10^{-9} & 6\times 10^{-8} \\
7\times 10^{-7} & 8\times 10^{-6} & 2\times 10^{-4}
} .
\end{align}
Taken together with the weaker experimental limits on $u\to u' J$ we can ignore these couplings in practice.

From the point of view of the seesaw expansion, the quark-flavor changing couplings are actually the dominant Majoron couplings. They only decouple as $\log (M_R)/f$, whereas all other couplings decouple at least as $1/f$. 
It was noted before that Goldstone bosons with effective diagonal couplings $J m_q \bar{q}\i\gamma_5 q$ yield flavor-changing quark couplings at one loop that depend logarithmically on the UV scale~\cite{Freytsis:2009ct,Batell:2009jf}, whereas an initial coupling $J W_{\mu\nu}\tilde W^{\mu\nu}$ does not have such a dependence~\cite{Izaguirre:2016dfi}.
In our case $J W_{\mu\nu}\tilde W^{\mu\nu}$ gives only a seesaw-suppressed contribution to $Jqq'$, and the $\log (M_R)$ terms originate from an effective coupling to Goldstone bosons, $JG^+G^-$.

\section{Phenomenology}
\label{sec:pheno}

Having obtained all Majoron couplings to leading order in the seesaw expansion we can discuss existing constraints and signatures.

\subsection{Light Majorons}
\label{sec:light_Majorons}

We start with the simplest case of a \emph{massless} Majoron, which most importantly gives a vanishing coupling to photons.
It proves convenient to phrase our discussion in terms of the dimensionless parameters
\begin{align}
K_{\alpha\beta} \equiv \frac{(M_D M_D^\dagger)_{\alpha\beta}}{v f} \,,
\end{align}
as they capture the Majoron couplings in most cases~\cite{Garcia-Cely:2017oco}.

The off-diagonal entries of $M_D M_D^\dagger$ are directly constrained by the lepton-flavor-violating (LFV) decays $\ell \to \ell' J$~\cite{Pilaftsis:1993af,Feng:1997tn,Hirsch:2009ee}.
For $m_{\ell'} \ll m_\ell$, the partial widths read
\begin{align}
\frac{\Gamma (\ell \to \ell' J)}{\Gamma (\ell \to \ell' \nu_\ell \bar\nu_{\ell'})} \simeq \frac{3}{16 \pi^2}\frac{v^2}{m_\ell^2}\, |K_{\ell\ell'}|^2
\end{align}
and involve a left-handed final-state lepton, leading to an anisotropic decay~\cite{Garcia-Cely:2017oco}.
The constraints in the tau sector  are  $\Br (\tau \to \ell J)<\mathcal{O}(10^{-3})$~\cite{Albrecht:1995ht} and lead to~\cite{Garcia-Cely:2017oco}
\begin{align}
|K_{\tau e}|< 6\times 10^{-3}\,, &&
|K_{\tau \mu}| < 9\times 10^{-3}\,,
\end{align}
which can be improved by Belle and Belle-II~\cite{Heeck:2016xkh,Heeck:2017xmg,Yoshinobu:2017jti}.
In the muon sector, the best constraints on a Majoron with anisotropic emission come from $\mu\to e J$~\cite{Bayes:2014lxz} (to be improved with Mu3e~\cite{Perrevoort:2018ttp}) and $\mu\to e J \gamma$~\cite{Goldman:1987hy}. The latter is also sensitive to $m_J=0$ and provides a limit
\begin{align}
|K_{\mu e}|  < 10^{-5}\,.
\end{align}
The \emph{diagonal} couplings $K_{\ell\ell}$ of a massless Majoron are constrained by astrophysics, stellar cooling in particular, and imply
\begin{align}
|K_{ee}-K_{\mu\mu} -K_{\tau\tau}| &< 2\times 10^{-5} \,, \\
 \tr (K) &< 5\times 10^{-6}\,,
\end{align}
from the electron~\cite{Raffelt:1994ry} and nucleon coupling~\cite{Keil:1996ju}, respectively.
The bound on the trace $\tr (K)$ is particularly powerful since it provides upper bounds on all entries of the positive-definite $K$~\cite{Garcia-Cely:2017oco} by means of the inequality of Eq.~\eqref{eq:inequality}. This then puts an upper bound on the Majoron coupling to muons and taus, which is far better than any direct bound on these couplings. It also ensures that rare decays such as $K\to \pi J$ and $Z\to \gamma J$~\cite{Bauer:2017ris,Craig:2018kne,Bauer:2018uxu,Alonso-Alvarez:2018irt} are unobservably suppressed for a massless Majoron.
Two-loop couplings are hence irrelevant for massless Majorons.

Overall we see that a massless Majoron gives seesaw-parameter constraints of order $M_D M_D^\dagger/(v f) \lesssim 10^{-5}$--$10^{-6}$. While this is far off the ``natural'' value $M_D M_D^\dagger/(v f)\sim M_\nu/v \sim 10^{-13}$, it can be realized by assuming certain matrix structures in $M_D$ that suppress $M_\nu \simeq -M_D M_R^{-1} M_D^T$ but not $M_D M_D^\dagger$, to be discussed in more detail in Sec.~\ref{sec:comparison_to_seesaw}. As we have seen, the relevant couplings of a massless Majoron are those to nucleons and electrons, but even $\mu\to e J$ could be observable.

\begin{figure}[t]
	\includegraphics[width=0.48\textwidth]{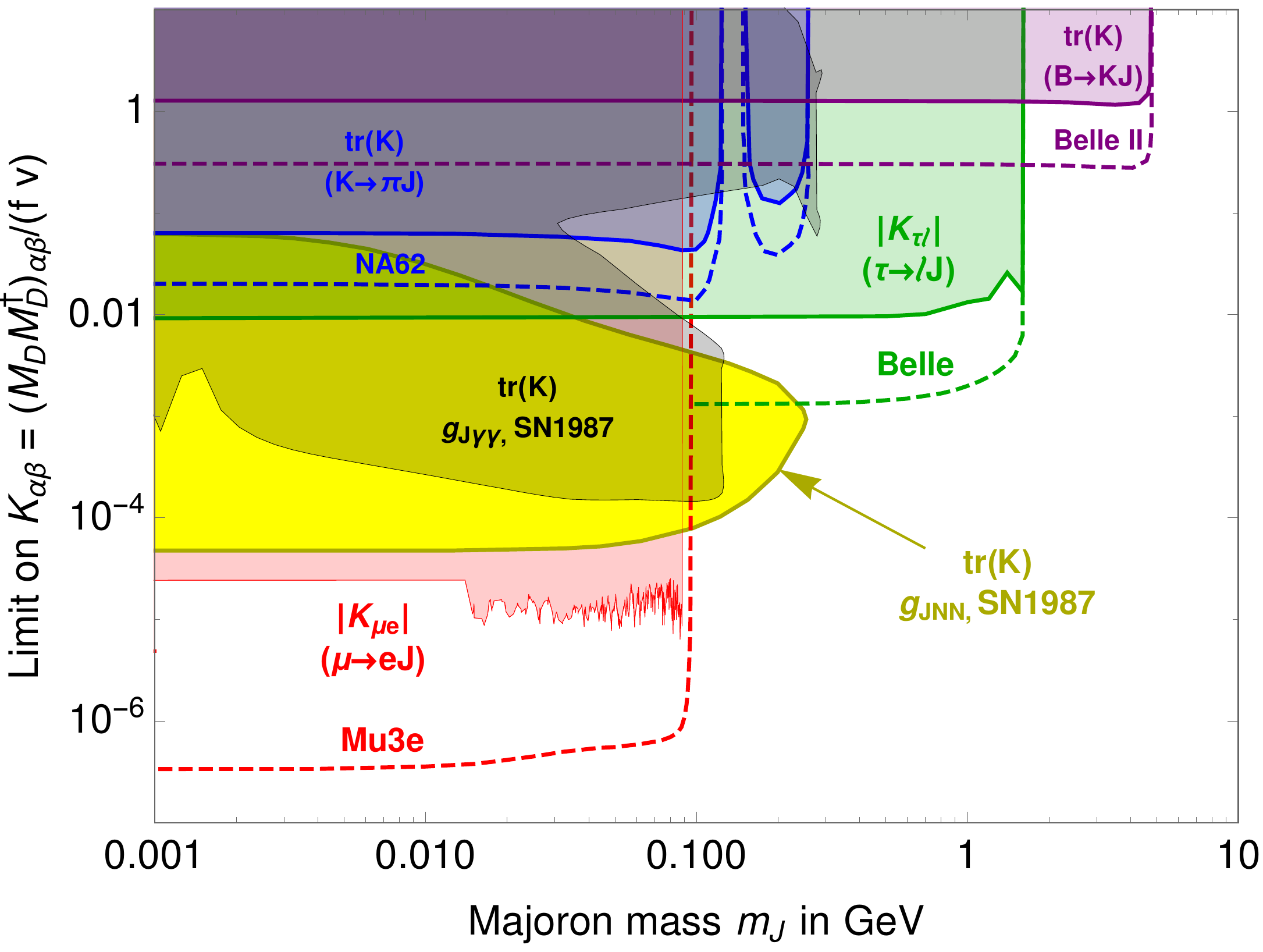}
	\caption{
		Upper limits on combinations of $K_{\alpha\beta}= (M_D M_D^\dagger)_{\alpha\beta}/(v f) $ for Majoron masses above MeV. The shaded regions exclude $|K_{\alpha\beta}|$ or $\tr (K)$ by non-observed rare decays, the dashed lines show the potential future reach; see text for details.
		$K\to \pi J$ and $B\to KJ$ further scale with $\log (M_R/m_W)$, which has been set to $\id$ here.
		The black region is a very naive estimate of SN1987 constraints on the di-photon coupling, setting for simplicity $K_{\ell\ell}=\tr(K)/3$.
		The yellow region is the SN1987 constraint on the $JNN$ coupling~\cite{Lee:2018lcj}.
		The off-diagonal entries have to satisfy $|K_{\alpha\beta} | < \tr(K)/2$; see Eq.~\eqref{eq:inequality}.}
	\label{fig:light_Majoron_constraints}
\end{figure}

The phenomenology becomes more interesting for non-zero Majoron mass, specifically values above $\sim\unit[10]{keV}$ in order to kinematically evade the stellar cooling constraints. This is shown in Fig.~\ref{fig:light_Majoron_constraints} for Majoron masses above \unit[1]{MeV}. In addition to the one-loop lepton-flavor-violating decays that probe $K_{\alpha\beta}$ we now also have relevant constraints from the two-loop quark-flavor-violating decays, especially $K\to \pi J$ and $B\to K J$ (limits and future sensitivity taken from Ref.~\cite{Gavela:2019wzg}, based on Refs.~\cite{Artamonov:2008qb,Adler:2004hp,Izaguirre:2016dfi}). 
These decays probe the quantity $\tr (M_D \log (M_R/m_W) M_D^\dagger)$, but to simplify comparison with the other limits we set $\log (M_R/m_W) = \id$ to obtain a limit on $\tr (K)$. It should be kept in mind, however, that a larger log enhancement can make these rare processes even more relevant.
Also potentially relevant are the two-loop Majoron couplings to photons via the effective coupling $g_{J\gamma\gamma}$ from Sec.~\ref{sec:diphoton}. Astrophysical limits on this coupling are extremely strong for $m_J \lesssim \unit[100]{MeV}$~\cite{Jaeckel:2017tud}; since $g_{J\gamma\gamma}$ is $m_J^2$ suppressed for $m_J \lesssim \unit{MeV} $, the region where the photon coupling is important is between MeV and $\unit[100]{MeV}$. This is unfortunately precisely the mass region where the light quarks that run in the $J$-$\gamma$-$\gamma$ loops should be replaced by hadrons; as a very naive way to incorporate this we simply set $m_u = m_d = m_\pi$ and $m_s = m_K$ in Eq.~\eqref{eq:JgammagammaSet1}. The $g_{J\gamma\gamma}$ limit in Fig.~\ref{fig:light_Majoron_constraints} should therefore not be taken too seriously. 
In light of these uncertainties we do not discuss how the $g_{J\gamma\gamma}$ coupling depends on the various diagonal $K_{\ell\ell}$, but rather set them all equal to $\tr(K)/3$ to allow a comparison to the other limits. It is clearly possible to suppress $g_{J\gamma\gamma}$ significantly in the region of interest by choosing hierarchical $K_{\ell\ell}$, as shown in Fig.~\ref{fig:Jgammagamma_amplitude}.

Also illustrated in Fig.~\ref{fig:light_Majoron_constraints} are SN1987 constraints on the Majoron--\emph{nucleon} coupling, Eq.~\eqref{eq:Jpp}, adopted from Ref.~\cite{Lee:2018lcj}, which reach up to $m_J\sim \unit[250]{MeV}$ and constrain $\tr(K)$ between $5\times 10^{-5}$ and $0.06$.

As can be appreciated from Fig.~\ref{fig:light_Majoron_constraints}, even the strong astrophysical constraints on Majorons do not rule out flavor-violating rare decays, with significant experimental progress expected in the near future.
Even the two-loop suppressed $d\to d' J$ decays provide meaningful constraints.

$\mu\to e J$ is well constrained already, and we expect it to eventually become the most sensitive probe of the $K$ matrix entries for $m_J < m_\mu$, even beating out stellar cooling limits. $\tau\to \ell J$, on the other hand, is mainly relevant for $m_J$ between $\sim\unit[100]{MeV}$ and $m_\tau$. For smaller $m_J$ the flavor-\emph{conserving} constraints on $K$ from $g_{J\gamma\gamma}$ and $g_{JNN}$ become stronger, which suppresses the LFV modes via the inequality of Eq.~\eqref{eq:inequality}.
For $m_J > m_\tau$, the main rare decays are $B\to K J$ and $Z\to J\gamma$~\cite{Brivio:2017ije}, the former is shown in Fig.~\ref{fig:light_Majoron_constraints}, and the latter gives irrelevant constraints on the $K$ entries of order $10^4$.

\subsection{Comparison with seesaw observables}
\label{sec:comparison_to_seesaw}

So far we have discussed the interactions of the Majoron, but, of course, the right-handed neutrinos $N_R$ also mediate non-Majoron processes, discussed at length in the literature~\cite{Broncano:2002rw,Broncano:2003fq,Broncano:2004tz,Cirigliano:2005ck,Abada:2007ux,Gavela:2009cd,Coy:2018bxr}. Assuming again that the $N_R$ are heavy enough to be integrated out, the relevant dimension-six operators involving SM fields all depend on the matrices
\begin{align}
M_D M_R^{-2} M_D^\dagger \,\, \text{ and }\,\, M_D M_R^{-2}\log\left( M_R/m_W\right) M_D^\dagger \,,
\label{eq:seesaw_matrices}
\end{align}
which drive LFV processes such as $\ell \to \ell' \gamma$ as well as lepton-universality violating effects such as $\Gamma(Z\to \ell\bar{\ell})/\Gamma(Z\to \ell'\bar{\ell}')$, recently discussed thoroughly in Ref.~\cite{Coy:2018bxr}.
In comparison, we have seen above that all Majoron operators depend on the matrices
\begin{align}
\frac{M_D M_D^\dagger}{f} \,\, \text{ and }\,\, \frac{M_D \log\left( M_R/m_W\right) M_D^\dagger}{f} \,.
\label{eq:Majoron_matrices}
\end{align}
For $f\sim M_R$, this makes the Majoron operators potentially dominant, while $f\gg M_R$ suppresses them to an arbitrary degree~\cite{Garcia-Cely:2017oco}. To properly compare Majoron and non-Majoron processes it is necessary to pick a structure for $M_D$, which is guided by our experimental reach.

\begin{figure*}[t]
	\includegraphics[width=0.48\textwidth]{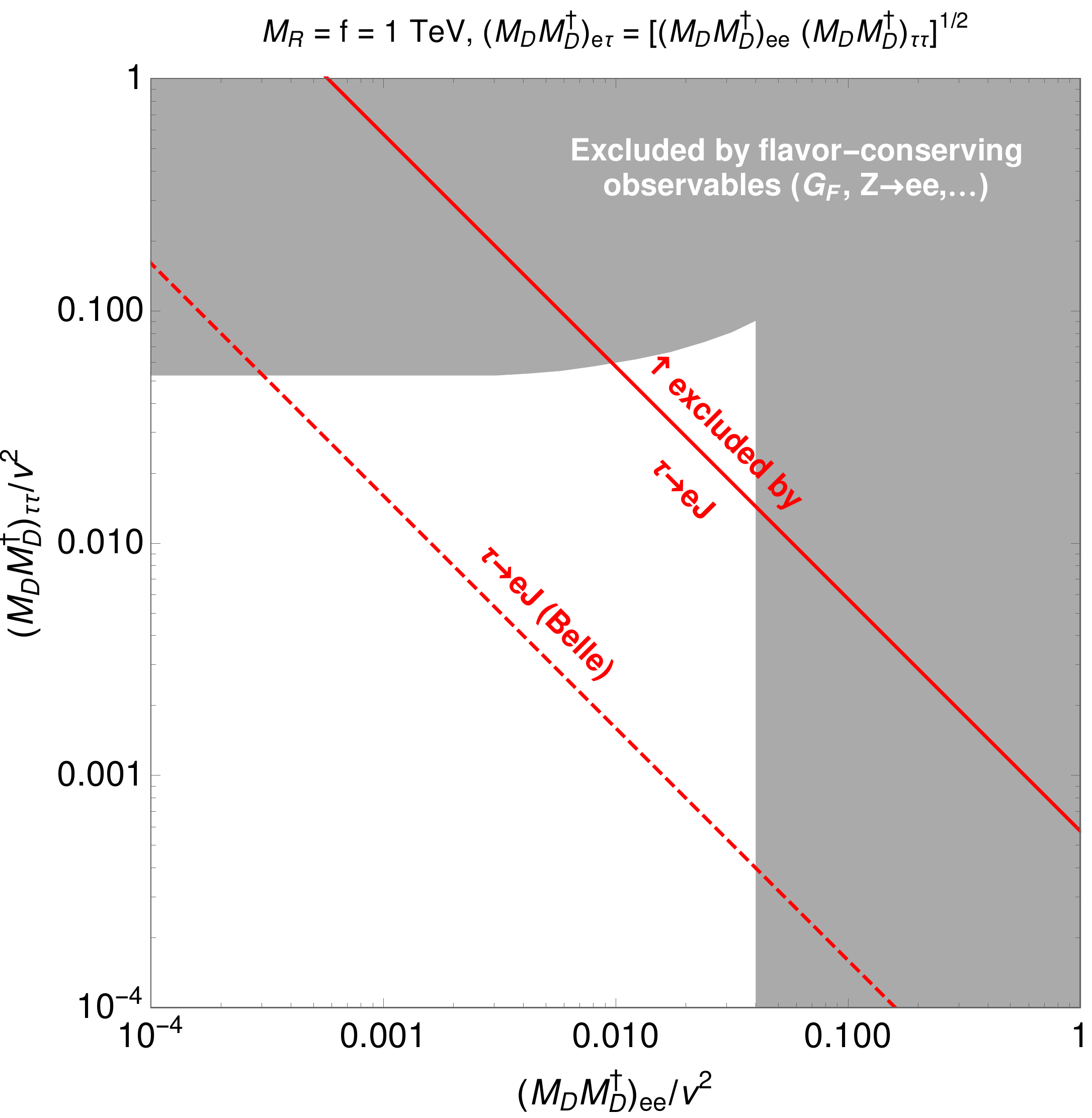}\hspace{2ex}
	\includegraphics[width=0.48\textwidth]{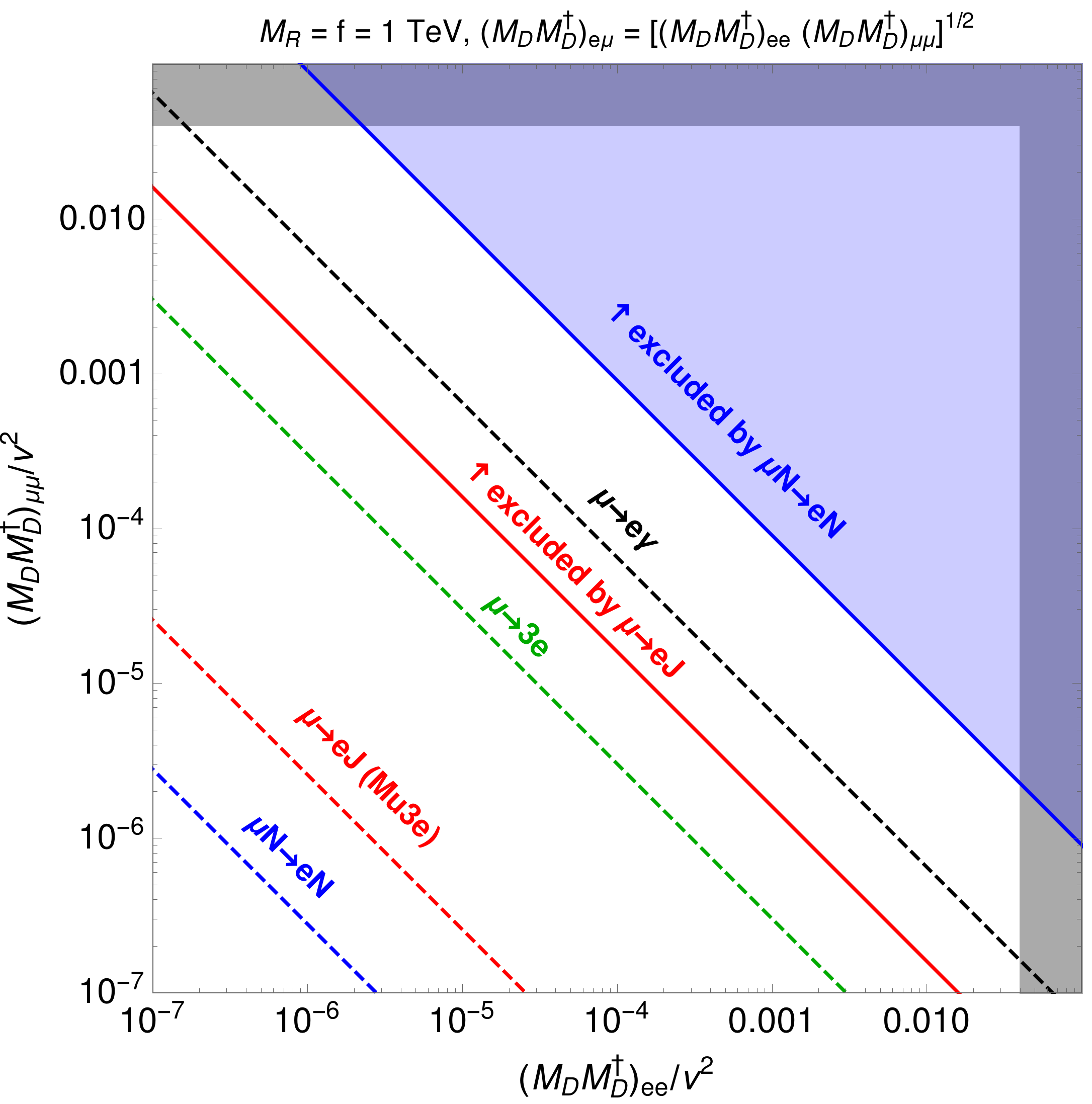}
	\caption{
		Limits on $(M_D M_D^\dagger)_{\alpha\beta} / v^2$ from flavor-conserving non-Majoron  (gray region) and flavor-violating observables (colored lines), dashed lines denoting future reach, adopted from Ref.~\cite{Coy:2018bxr}. We set $M_R/\unit{TeV}=\id$, $f=\unit[1]{TeV}$, and assume a matrix texture that gives $(M_D M_D^\dagger)_{\alpha\beta}= \sqrt{(M_D M_D^\dagger)_{\alpha\alpha}(M_D M_D^\dagger)_{\beta\beta}}$; see text for details. The Majoron mass is assumed to be between $\unit[100]{MeV}$ and GeV in the left figure and between MeV and $\unit[100]{MeV}$ in the right figure.}
	\label{fig:seesaw_limits}
\end{figure*}

As we have seen above, even future limits on Majoron production cannot reach the natural seesaw scale $M_R\sim\unit[10^{14}]{GeV}$, and the same is true for other $N_R$-mediated processes~\cite{Coy:2018bxr}. By no means does this preclude observable effects, since it is possible to use the matrix structure of $M_D$ to suppress $M_\nu \simeq -M_D M_R^{-1} M_D^T$ while keeping $M_D M_D^\dagger$ large~\cite{Buchmuller:1990du,Buchmuller:1991tu}, potentially realizing a lepton-number symmetry~\cite{Kersten:2007vk} as in the inverse seesaw~\cite{Wyler:1982dd,Mohapatra:1986bd,GonzalezGarcia:1988rw}. Following Refs.~\cite{Ingelman:1993ve,Coy:2018bxr} we can solve $M_\nu =0$ for $M_D$, which then requires only tiny perturbations $\delta M_D$ to produce the observable neutrino masses via
\begin{align}
M_\nu \simeq - \delta M_D M_R^{-1} M_D^T- M_D M_R^{-1} (\delta M_D)^T \,.
\end{align}
The key observation is that $\delta M_D$ is negligible in $M_D M_D^\dagger$. $M_\nu =0$ requires the low-scale seesaw structure
\begin{align}
M_D = v \matrixx{\xi_e \\ \xi_\mu \\ \xi_\tau} \matrixx{1 && z \sqrt{\frac{m_5}{m_4}} && \pm \i \sqrt{1+z^2}\sqrt{\frac{m_6}{m_4}}} ,
\label{eq:low_scale_texture}
\end{align}
with complex $z$ and real $\xi_{\alpha}$ without loss of generality~\cite{Coy:2018bxr}.\footnote{To ensure $M_\nu =0$ to all orders in the seesaw expansion and at loop level one has to further impose either $m_5=m_4$ and $z=\pm \i$ or $m_6=m_4$ and $z=0$, both of which correspond to a conserved lepton number~\cite{Kersten:2007vk,Moffat:2017feq}.} This product structure of $M_D$ implies
\begin{align}
(M_D \,b(M_R) M_D^\dagger)_{\alpha\beta} \propto \xi_\alpha \xi_\beta \,,
\end{align}
for any function $b$. The off-diagonal entries are then \emph{real} and entirely determined by the diagonal ones:
\begin{align}
&(M_D \,b(M_R) M_D^\dagger)_{\alpha\beta} \label{eq:equality}\\
&\qquad = \sqrt{(M_D \,b(M_R) M_D^\dagger)_{\alpha\alpha} (M_D \,b(M_R) M_D^\dagger)_{\beta\beta}} \,.\nonumber
\end{align}
Notice that this violates the strict inequality derived earlier in Eq.~\eqref{eq:inequality}, which assumed \emph{non}-vanishing neutrino masses.

Equation~\eqref{eq:equality} drastically simplifies the discussion of the seesaw parameter space, seeing as all observables now only depend on the three real diagonal entries of $M_D\, b(M_R) M_D^\dagger$ instead of the nine parameters it could contain in general.
Furthermore, all Majoron and non-Majoron parameter matrices have the same flavor structure and only differ in their absolute magnitude.
Equation~\eqref{eq:equality} also shows that any low-scale seesaw texture automatically \emph{maximizes} the off-diagonal flavor-violating entries of the relevant coupling matrices (Eq.~\eqref{eq:seesaw_matrices} or Eq.~\eqref{eq:Majoron_matrices}).

We compare Majoron and non-Majoron limits in Fig.~\ref{fig:seesaw_limits}, the latter adopted from Ref.~\cite{Coy:2018bxr}. We set $M_R/\unit{TeV} =\id$ as well as $f=\unit[1]{TeV}$ and stress again that the Majoron limits can be suppressed arbitrarily by increasing~$f$.
Figure~\ref{fig:seesaw_limits} (left) shows the $(M_D M_D^\dagger)_{ee}$ vs.~$(M_D M_D^\dagger)_{\tau\tau}$ parameter space, setting $(M_D M_D^\dagger)_{\mu\mu}=0$. As already noted in Ref.~\cite{Coy:2018bxr}, the non-Majoron limits are completely dominated by flavor-\emph{conserving} observables such as $Z\to e^+e^-$ and other electroweak precision data. LFV such as $\tau\to e\gamma$ and $\tau\to eee$ reside far in the already excluded region and even future improvements, e.g.~in Belle-II, do not reach the allowed parameter space. In comparison, Majorons with masses between $\sim\unit[100]{MeV}$ and $m_\tau$ do give relevant constraints from $\tau\to e J$ and will probe significantly more parameter space with upcoming Belle and Belle-II analyses~\cite{Yoshinobu:2017jti}. Standard LFV in the $\tau e$ (and $\tau\mu$ in complete analogy) sector are hence doomed to be unobservable in the seesaw model, but the Majoron LFV channels $\tau \to \ell J$ could be observable and deserve more experimental attention.

Figure~\ref{fig:seesaw_limits} (right) shows the $(M_D M_D^\dagger)_{ee}$ vs.~$(M_D M_D^\dagger)_{\mu\mu}$ parameter space, setting $(M_D M_D^\dagger)_{\tau\tau}=0$. Standard LFV, currently dominated by $\mu$ conversion in nuclei~\cite{Bertl:2006up}, provides important constraints on the parameter space, and all future $\mu e$ LFV will probe uncharted terrain. For $m_J < m_\mu$, the Majoron channel $\mu \to e J$ already sets  better limits than $\mu N \to e N$ and can continue to dominate over $\mu\to e \gamma$ and $\mu\to 3e$ in the future. Ultimately, $\mu N \to e N$ conversion in Mu2e~\cite{Abrams:2012er} and COMET~\cite{Adamov:2018vin} has the best future reach.

\subsection{Majoron dark matter}
\label{sec:Majoron_dark_matter}

Returning to the ``standard'' high-scale seesaw scenario with huge hierarchy $v\ll f$ it is clear that a massive Majoron can be long lived even on cosmological scales, e.g.~from Eq.~\eqref{eq:J_to_nunu},
\begin{align}
\Gamma (J\to \nu\nu) \sim \frac{1}{\unit[400]{Gyr}}\left(\frac{m_J}{\unit{MeV}}\right)\left(\frac{\unit[10^9]{GeV}}{f}\right)^2 .
\end{align}
In this region of parameter space Majorons can form DM~\cite{Rothstein:1992rh,Berezinsky:1993fm,Lattanzi:2007ux,Bazzocchi:2008fh,Frigerio:2011in,Lattanzi:2013uza,Queiroz:2014yna, Wang:2016vfj}, with a production mechanism that can be unrelated to the small decay couplings~\cite{Frigerio:2011in,Garcia-Cely:2017oco,Heeck:2017xbu,Boulebnane:2017fxw}.

The defining signature of Majoron DM is a flux of neutrinos  from DM decay with $E_\nu \simeq m_J/2$ and a known flavor composition~\cite{Garcia-Cely:2017oco}. For $m_J \gtrsim \unit{MeV}$ these neutrino lines could potentially be  observable via charged-current processes in detectors such as Borexino or Super-Kamiokande~\cite{Garcia-Cely:2017oco}, while lower masses are more difficult to probe~\cite{McKeen:2018xyz}.

The loop-level couplings generate the much more constrained decays $J \to f\bar f', \gamma\gamma$, which, however, depend on the matrix $K$ and are hence \emph{complementary} to the neutrino signature, as discussed in detail in Ref.~\cite{Garcia-Cely:2017oco}. This analysis used only the Set~II of diagrams to calculate $J\to\gamma\gamma$, namely the expression proportional to $\tr (K)$, Eq.~\eqref{eq:JgammagammaSet1}. The new full expression presented in Sec.~\ref{sec:diphoton} leads in general to a \emph{suppression} of the diphoton rate and a more involved dependence on the $K$ matrix entries (Fig.~\ref{fig:Jgammagamma_amplitude}).
We omit a full recasting of existing DM$\to\gamma\gamma$ limits onto our $J\to\gamma\gamma$ expression since it is not very illuminating, but stress that this diphoton suppression makes the neutrino modes even more dominant.

\section{Conclusion}
\label{sec:conclusion}

Majorons, the Goldstone bosons of spontaneously broken lepton number, were proposed in the early 1980s in models for Majorana neutrino masses. Since then experiments have indeed found evidence for non-zero neutrino masses, although it is not yet clear  whether they are of Majorana type. With the motivation for Majorons as strong as ever, we have set out in this article to complete the program that was started almost 40 years ago and calculate all Majoron couplings to SM particles. The couplings to neutrinos (tree level) as well as charged leptons and diagonal quarks (one loop) were known previously.  Here we presented the two-loop couplings to gauge bosons ($J\gamma\gamma$, $J\gamma Z$, $JZZ$, $JWW$, $Jgg$) and flavor-changing quarks ($Jdd'$, $Juu'$). Phenomenologically relevant of these are currently only the Majoron coupling to photons as well as the $Jdd'$ couplings behind the rare decays $K\to \pi J$ and $B\to K J$.

Standard seesaw effects in an EFT approach are encoded in the matrix $M_D M_R^{-2} M_D^\dagger$, which drives, for example, $\ell\to \ell' \gamma$. Majoron couplings, on the other hand, depend on the matrix $M_D M_D^\dagger/f$, which is parametrically larger in the seesaw limit and can indeed give better constraints in parts of the parameter space. For example, while $\tau \to \ell \gamma$ and other $\tau$ LFV are unlikely to be observable in the seesaw model, $\tau\to \ell J$ can be observably large and deserves more experimental attention.

The singlet Majoron model together with the coupling texture of Eq.~\eqref{eq:low_scale_texture} implied by low-scale seesaw is a very minimal UV-complete realization of an axion-like particle and thus a well-defined benchmark model.
The dominant theoretical challenge not addressed here is the replacement of quarks by hadrons in loops, which we leave for future work.
We expect future studies to elucidate additional aspects of this model, in particular when the Majoron is used as a portal to dark matter.

\section*{Acknowledgements}
JH would like to thank Arvind Rajaraman, Raghuveer Garani, and C{\'e}dric Weiland for useful discussions.  HHP would like to thank Wolfgang Altmannshofer, Michael Dine, and Stefano Profumo for useful discussions.  JH is supported, in part, by the National Science Foundation under Grant No.~PHY-1620638, and by a Feodor Lynen Research Fellowship of the Alexander von Humboldt Foundation.
The work of JH was performed in part at the Aspen Center for Physics, which is supported by the National Science Foundation under Grant No.~PHY-1607611.
HHP is partly supported by U.S.~Department of Energy Grant No.~de-sc0010107.

\bibliographystyle{utcaps_mod}
\bibliography{BIB}

\end{document}